\documentstyle{article}
\def\bi{\begin{list}{$\bullet$}{\topsep0pt \parsep=\parskip
\itemsep=0pt}}
\def\ei{\end{list}}
\def\sec#1\par{\let\Large=\relax\section{\uppercase{\bf#1}}}
\def\subsec#1\par{{\let\large=\relax\subsection{#1}}\par\noindent}
\def\[{\begin{equation}}
\def\]{\end{equation}}
\def\mede#1{\left\langle#1\right\rangle}
\begin{document}
\thispagestyle{plain}
\begin{center}
{\svtnrm  Fast low  fidelity microsimulation  of  vehicle  traffic  on
supercomputers }\\[2mm]
Kai Nagel\footnote{permanent address: ZPR c/o Math.\ Inst., Univ.,
Albertus-Magnus-Platz, 50923~K\"oln, Germany}
\\[2mm]
Los Alamos National Laboratory, A-DO/SA, MS 997, Los Alamos, NM 87545, U.S.A.\\
and\\
Santa Fe Institute, 1660 Old Pecos Trail, Santa Fe, NM 87501, U.S.A.\\[2mm]
\today
\end{center}
\vskip5mm
\noindent{\bf Abstract: }
A set of very simple rules for driving behavior used to simulate
roadway traffic gives realistic results.  Because of its
simplicity, it is easy to implement the model on supercomputers
(vectorizing and parallel), where we have achieved real time limits of
more than 4~million~kilometers (or more than 53~million vehicle
sec/sec).  The model can be used for applications where both high
simulation speed and individual vehicle resolution are needed.  We use
the model for extended statistical analysis to gain insight into
traffic phenomena near capacity, and we discuss that this model is a
good candidate for network routing applications.

\sec Introduction

The task of optimally routing a multitude of vehicles through a road
network is extremely complex.  Important reasons for this are the
non-linear cost function (the delay due to high traffic is not
proportional to the traffic density), the stochasticity and
time-dependency of the traffic demand, and the nonlinear and
stochastic properties of traffic dynamics.  Therefore, the problem is
intractable by the so far known methods of Combinatorial Optimization
used in Operations Research, which means that no mathematical
technique is known for finding a globally optimal assignment of
traffic demand to a route network under realistic conditions.

This led us to the idea of using a very fast simulation model for
developing and testing mathematically based optimization techniques.
In order to simulate the effect of individual routing, the simulation
model had to be microscopic, i.e.\ modeling each vehicle as an
individual entity with the possibility of a different track plan for
each vehicle.  To get high computational speed, we formulated the
traffic dynamics with as few rules as possible.  Due to this
simplicity the model can in addition be efficiently installed on many
kinds of supercomputers~\cite{NaS93}.

Besides this, the model proves to be unexpectedly
realistic~\cite{NaS92} and therefore interesting in its own right.  In
consequence, it has become the object of ongoing research using
analytical and computational techniques (see~\cite{ScS93,NaH93};
further publications are in preparation).

An interesting phenomenon in real traffic reproduced by our model is
the formation of start-stop-waves in the vicinity of the regime of
maximum throughput.  These start-stop-wave instabilities are the
phenomenon which restricts the vehicle throughput although the waves
itself appear already far below the ``critical'' traffic density.
Theoretical work based on fluiddynamical analogies~\cite{LW} points
out similarities to a transition from laminar to turbulent
flow~\cite{Kue,Persaud}, and experimental evidence confirms the
hypothesis of nontrivial behavior and large
fluctuations~\cite{ZKB,Musha}.

The outline of this paper is as follows: At first, in section~2, we
recall the definition of our traffic model.  Section~3 compares
simulation output with real data.  In section~4, we give an overview
over implementations on different supercomputer architectures and
discuss performance results in view of traffic applications.
Section~5 contains several examples of how the model may be used as a
tool to facilitate a further understanding of traffic patterns near
capacity.  We finish with a short summary/conclusion.

\sec Description of the model

The principal idea of our approach is to describe the behavior of
vehicles by a set of rules which are as simple as possible.  This
obviously implies a tradeoff between realism and simulation speed.  We
describe in the following the set of rules for a simple single-lane
traffic flow model; multi-lane modeling will be the subject of later
publications.

The model is defined on a one-dimensional array, cutting the street
into boxes of length $\epsilon \approx 7.5\,m\ (\approx 23~feet)$.
This is the length one vehicle approximately occupies in a congestion.
Each box is either empty, or occupied by one vehicle.  Since positions
are limited to integer array positions, velocities can also only be
integers, with a value between $0$ and $v_{max}$.  When a vehicle for
instance has the velocity~3, it will jump 3~boxes forwards in a time
step.

We use $v_{max} = 5$.  In fact, the phenomenology of the model (i.e.\
the formation of start-stop-waves) is independent of the choice of
$v_{max}$, but the form of the resulting fundamental diagram
(throughput vs.\ density, see below) becomes unrealistic for values
larger than~5 or smaller than about~3~\cite{NaS93}.

Before the propagation of a car, its velocity is adapted according to
its specific situation:\bi

\item
{\it slowing down:} If the next car ahead is too close ($gap \le v-1$,
where $gap$ is the number of empty boxes between a car and the next
one ahead), then the velocity is reduced: $v \to gap$.

\item
{\it acceleration:} If, however, the gap is large enough ($gap \ge
v+1$) {\it and\/} if the velocity is smaller than the maximum velocity
($v
\le v_{max} - 1$), then the velocity is increased by one: $v \to
v+1$.

\ei
These rules use as input only ``old'' information which is not changed
or generated during the update.  Therefore, this deterministic part of
the velocity update may be performed ``in parallel'', i.e.,
simultaneously on all vehicles.

In order to take into account the natural fluctuations of driving
behavior, we add a\bi

\item
{\it randomization:} When velocity is 1 or larger ($v \ge 1$), then
with probability $p=0.5$ the velocity of each vehicle may in addition
to the above rules be reduced by one: $v \to v - \gamma(p)$ , where
$\gamma(p)$ is one with probability $p$ and zero else.

\ei
The randomization introduces speed fluctuations for free driving,
overreactions at slowing down, and retardations during acceleration.

Note that for the whole velocity update of one vehicle we use only
{\it position\/} information of the predecessor---this is in contrast
to what some car following theories suggest~\cite{yyyy}. Nevertheless,
this leads to realistic results (see later).

As already mentioned above, after the velocity update each vehicle is
advanced $v$ boxes to the right (``{\it propagation step\/}'').  Since
this step, as well as the randomization, can again be performed in
parallel, the whole update consists of parallel rules, a fact which
simplifies enormously the issue of efficiently programming on
supercomputers.

When taking into account randomization, the average free speed of a
vehicle in our model is $4.5 \times 7.5~m/\Delta t$, where $\Delta t$
is the model time step.  This should be equal to about $112~km/h$
$70~mph$), which results in $\Delta t \approx 1~sec$.

\sec Phenomenology of the model

Fig.~\ref{evol.std} shows a typical evolution of the system from
random initial conditions.  The horizontal direction is the space
direction; time is pointing downwards.  Each black pixel corresponds
to a vehicle moving from left to right; each horizontal line therefore
shows the configuration of the simulated road segment at a different
time step.  In fact, the plot is similar to the well-known
space-time-diagrams from aerial photography~\cite{Treiterer}, except
that time and space axes are exchanged, and that one has to connect
all pixels which represent one car in order to get its trajectory.

One observes in the figure that the details of the random initial
conditions quickly become irrelevant and that the system's appearance
is dominated by traffic jams of high vehicle density (black),
separated by zones of free traffic.  Jams form and dissolve at
arbitrary times and positions, as can be seen at some of the smaller
jams.

As a next step of our investigation, we measured the relation between
throughput and density for our model.  Averages over $T =
200$~time steps, measured locally at one site, result in the
scatter-plot of Fig.~\ref{fdiag}.  (To be specific, density~$\rho$ and
throughput~$q$ at position~$x_0$ are here defined by
\[
\rho_{T} := {1 \over T} \, \sum_{t=t_0}^{t_0+T-1} \delta(x_0,t)
\qquad\hbox{ and }\qquad
q_T := {n \over T} \ ,
\]
where $\delta(x_0,t)$ is one if there is a vehicle at position~$x_0$
at time~$t$ and zero else, and $n$ is the number of vehicles which
passed at position~$x_0$ during the time interval~$T$.)

Our simple model reproduces the linear relation between flow and
density for low traffic correctly.  Furthermore, near the density
corresponding to maximum capacity, it shows the strong fluctuations in
the capacity values~\cite{Kue,ZKB}, and these fluctuations as well as
the average throughput decrease approximately linearly with density at
higher densities.  The density corresponding to maximum flow is
somewhat low compared with reality, but this is corrected when using a
model for more than one lane and with slower vehicle types
(preliminary data (M.~Rickert); final results together with a detailed
description of the multi-lane algorithm will be the subject of a
future publication).

\sec Computational issues

\subsec Coding

Two obvious ways to implement the described dynamics are (i) {\it
site-oriented}, and (ii)~{\it vehicle-oriented} (see~\cite{NaS93} for
a more detailed description).  Site-oriented represents a street by an
array $(v_i)_i$ of integers with state values between -1 and
$v_{max}$. A value of $v_i = -1$ means that there is no car on
site~$i$, whereas a value $v_i$ between 0 and $v_{max}$ denotes a car
with speed~$v_i$ at site~$i$.  The state of each site can then be
updated as a function of the states of its neighbors.  The principal
disadvantage of this technique is that one needs as much computer time
for the empty sites as for the occupied ones; but the advantage is
that one can employ sophisticated computing techniques
(e.g.~single-bit coding) known from advanced cellular automata (CA)
simulations~\cite{Stauffer}, which yields very high efficiency
especially on SIMD architectures.  (An early but elaborated version of
the site-oriented approach is~\cite{Cremer}, which even uses
single-bit coding.  But it uses the technique in a way which is not
vectorizable and can therefore not use the power of many
supercomputers.)

In contrast, the {\em vehicle-oriented\/} approach represents a street
as a list of pairs  {\em (position  of vehicle,  speed  of  vehicle)}.
Obviously,   the   vehicle-oriented  approach   will   outperform  the
grid-oriented  one for very low vehicle densities,  but for  densities
corresponding  to capacity  flow  the  results depend on  the computer
architecture and  on the  system size.  In addition, this approach  is
difficult to handle efficiently as soon as vehicle passing is allowed.
This is  especially  true on parallel computers,  where for efficiency
neighboring vehicles should reside on the same node.

In addition, we used a third ``intermediate approach'', whose data
structure is in principle site-oriented, but where the update only
treats the ``interesting sites''.  CA-techniques are no longer
applicable, but at least for non-vectorizing computer architectures
the loss in performance is never more than a factor of four.  In
addition, treatment of multi-lane traffic is easiest with this
approach.

\subsec Computer architecture overview

For our comparisons, we used a SUN Sparc10 workstation, a net of these
workstations coupled by ethernet under PVM, a NEC-SX3/11, an Intel
iPSC/860-Hypercube with 32~nodes, a Thinking Machines CM--5 with
32~nodes and a Parsytec GCel-3 with 1024 nodes.

The SX3 is a very advanced example of the traditional vectorcomputers,
comparable to the Cray-line.  Its power mostly stems from a
combination of vectorization and pipelining, the former meaning that
data which lies in some regular way in memory can be treated without
loosing time for loading and storing, and the latter meaning that the
output from one operation can directly be feeded into some other unit
which performs the next operation, and which works simultaneously with
the first unit.  For practical purposes, these machines may be seen as
SIMD (single instruction multiple data) machines.  But these
architectures have reached physical limits.

Microprocessors like in workstations or in personal computers have
become more and more powerful and sophisticated during the recent
years.  Therefore, an obvious idea to obtain high performance is to
combine many of these microprocessors to so-called parallel computers.
All processors act largely independently (MIMD: Multipe Instruction
Multiple Data).  A standard means to exchange information between
these different processors is ``message passing'', where one processor
sends out a message to another, but the message is only received when
the receiving node explicitly issues a receive command.  These
message passing commands are added to standard Fortran or C.  The
Parsytec GCel-3 and the Intel iPSC/860 Hypercube are two examples of
this type, the first one being a massively parallel machine containing
1024~relatively slow processors, and the second one modestly parallel
with 32~workstation-like CPU's.

These architectures behave in many respects like coupled workstations;
and  it  is indeed possible to use coupled workstations  as a parallel
machine.  Software packages like PVM (`parallel virtual machine')%
\footnote{Persons having access to electronic mail can obtain
information  on PVM by sending email to {\tt netlib@ornl.gov} with the
line {\it  send  index  from  pvm\/} in  the  subject field.   } offer
message passing  routines  to be included  into standard  Fortran or C
programs, and these messages are  transmitted e.g.\  via  the standard
ethernet.  However,  besides the slow  speed of the standard  networks
compared to those of the dedicated parallel machines, they encounter a
more serious principal problem: Networks like Ethernet  or  even  FDDI
(optical  link)  support  only  one message  at a  time  on the  whole
network.   Adding  further  machines  to  a  network  {\it  reduces\/}
therefore the amount of  time each  machine can  use  the  network for
communication.   This is essentially different  for dedicated parallel
architectures, where adding further processors usually does not change
the bandwidth between two (neighboring) processors.

The last machine we used is a CM-5, which has not only a workstation
processor on each node, but in addition 4~vector units.  If one does
not use the vector units, the machine behaves essentially similar to
the iPSC/860; but using the vector nodes represents a combination of
the traditional vector machines and the new parallel machines.  Using
the vector nodes involves up to now the use of a data parallel
programming language (High Performance Fortran or C$*$).

\bigskip\subsec Computational speed

Table~1 gives computational results for the implementations we tested;
as already mentioned, further details of the implementations, how they
relate to the different machines, and more detailed  performance  data
are given in~\cite{NaS93}.

When comparing performance data, it  is necessary to  give the size of
the simulated system.  This becomes imperative for parallel computers,
because too small systems perform poorly because of the  communication
overhead.  All values of Table~1 have been obtained by  simulations of
systems  of  size~$L  = 10,000~km$  ($6,250~miles$, $1,333,333~sites$)
with    an    average    traffic    density    of   $13.4~vehicles/km$
($8.4~vehicles/mile$,  $134,000~vehicles$ in the  whole system).  This
is a system size which seems relevant for  applications.  Moreover, it
is a system size small enough to still fit  into  memory of our single
node machines,  but  which  is at the same  time  large enough to  run
relatively efficient  on our parallel machines.   Quantitatively, this
means  that  both  the  GCel  and  the  CM-5  were  operating at  40\%
efficiency.   Thus, a system  size twice as large would need less than
twice as much simulation time on these machines.

References in the literature sometimes give a ``real time limit'' as
measure of their model's performance, which then is the {\it
extrapolated\/} system size (or number of vehicles) where simulation
is as fast as reality.  As explained above, we found these values
practically useless in the area of parallel computing, except when
given in conjunction with the system size which has really been
simulated.

In consequence and in order to avoid confusion, our primary table
entries are the CPU times we needed on the different machines in order
to simulate the system as defined above.  For convenience, we
calculated the real time limits in $km$ and in $vehicle~sec/sec$ from
these values.  But it should be kept in mind that, if one really
simulates system sizes near 1~million~$km$ on the parallel machines,
he/she will find much higher real time limits for these system sizes
(e.g.\ 2~million~$km$ instead of 900,000~$km$ on the GCel.

Noteworthy features of the table are: (i)~The CA-like algorithm is far
superior over the  ``intermediate'' one  on  the vectorizing  machines
(NEC  SX-3/11  and  CM-5), slightly  faster on  the  workstation-based
architectures, and slightly slower on  the massively parallel Parsytec
GCel-3.    (ii)~Both  algorithms  can  take  good   advantage  of  the
parallelism.   (iii)~The multilane-version  (which is developed  from
the ``intermediate'' algorithm) is, on the workstation, only  a factor
of  about two slower than the corresponding  single-lane version.   We
are  therefore confident to reach, for a realistic network setup, real
time limits of,  say, 430,000~single lane kilometers ($5,700,000~veh\,
sec/sec$) on 1024~nodes of the Parsytec~GCel-3, or of 1,700,000~single
lane kilometers  ($23,000,000~veh\,sec/sec$) on  512~nodes  of  a CM-5
(even without using the vector nodes).

\bigskip\sec Traffic dynamics near capacity

So far, we have shown  or argued that, despite its simplicity, the
model shows realistic properties, and that  one  can use
the   approach  for  very  fast,  low  fidelity,  microscopic  traffic
modeling.  In the  following,  we will  explain how the method  may be
used  to gain additional insight into  traffic near capacity,  and how
the traffic dynamics changes with  changes  in the  parameters of  the
model.   Our main argument is that we,  with  a  modest amount of
computer resources,  can give  answers  to  statistical questions such  as
``What happens to  capacity when everybody were equipped with a cruise
control?''or ``What ingredients are  necessary to make  a  speed limit
enhance  throughput~\cite{ZKB}?''.  Any findings certainly have to  be
inspected  more  closely  either  by  real  measurements  or  by  more
extensive  models, but these more  extensive  resources could then  be
directed towards crucial questions pointed out by this model.

One unquestionable advantage of computer models over reality is that
statistical averages are easily available; and with our model running
on a supercomputer, high-quality quantitative answers to many
questions can be obtained in days or even hours.  One example for this
is a time-averaged fundamental diagram (throughput vs.\ density, see
Fig.~\ref{fdiag}).  For this, we simulate a closed system (i.e.\
traffic ``on a ring'') of size $L$, $L$ being the number of ``boxes''.
An average density $\overline \rho := N/L$ ($N$: number of vehicles)
is easily defined because the number of vehicles is conserved during a
simulation run, and other quantities such as average throughput~$q$ or
average velocity~$v$ can be obtained by mimicking reality, i.e.,
counting the number of vehicles passing at a certain site, and
averaging over their velocities.

We find that we need systems of a size of at least $L = 10^4$
(corresponding to 75~km, or 1000~vehicles near capacity density) in
order to prevent arbitrary finite size effects due to the ring
geometry, and at least $10^6$~iterations to obtain satisfying averages
(especially near capacity flow, where fluctuations are largest).  In
our fastest implementation (on a NEC-SX3/11 single node
vectorcomputer), one such run needs about 30~sec, and about 100~such
runs covering the whole range of densities are needed for a meaningful
fundamental diagram ($q$-$\rho$-curve).

The maximum~$q_{max}^\infty$ of this  curve is the maximum capacity of
our model  traffic: In the long  run, it is not  possible to  get more
vehicles over the  segment.  This is not  in contrast to  the capacity
drop issue~\cite{PeH}---it simply  means that although for short times
higher  capacities  are     possible (cf.~short-time   averages     in
Fig.~\ref{evol.std}),   they   cancel out   against  periods  of lower
capacity in the long run.  The  only reasonable definition of capacity
seems to  be the longterm average,  which certainly  is much easier to
obtain from a model than in reality.

A simple  explanation is found  when looking at the space-time-plot of
Fig.~\ref{evol.std},  which is in  fact showing  the  system  near its
``threshold density''  $\rho^*  :=  \rho(q_{max}) \approx   0.08$: One
observes  that  start-stop-waves are  {\it natural\/}  for  systems at
maximum  throughput,  and  these waves  only   vanish at densities and
throughputs  far below the  capacity situation.   A capacity drop  may
therefore be seen as the result of a start-stop-wave.

Results are different for short segments.  Simulations of short closed
systems (Fig.~\ref{fdiag}) give a  higher maximum throughput for these
systems,  indicating that  there   is  a  way  to  feed  {\it short\/}
bottlenecks in a  way that the bottleneck capacity  is higher than the
capacity  $q_{max}^\infty$ of long segments.  But  this feeding has to
be flexible: Constant feeding leads to a throughput {\it well below\/}
capacity~\cite{NaS92}.  The overall lesson from this  is that there is
a well defined capacity in a simulation  model; a higher throughput is
possible for short periods or on short segments; and the latter is not
accessible  for traffic in a   standard bottleneck situation, but only
when  additional vehicles  are  injected  into the  traffic  flow {\it
inside\/} the bottleneck  (as by entry-ramps inside construction  site
bottlenecks).  A detailed publication on these matters and on relating
them to length and time scales of the traffic jams is in preparation.

Similarly, one can quantify velocity fluctuations as a function of
density (Fig.~\ref{fdiag}).  Technically, we measured the root mean
square deviation of the local velocity from its mean $\sigma(v) :=
\mede{ \mede{v^2} - \mede{v}^2 }$ where $\mede{\ldots}$ is the mean
over   all  car  which  have   passed  during   the measurement.   The
measurements show that  fluctuations are  low   for the  free  traffic
regime (apart from  the noise level  introduced by the randomization).
They increase rather rapidly near capacity  and reach their maximum at
a density {\it above\/} capacity.  Then they decrease, down to zero at
$\rho =  1$.  This means  that  the observations in~\cite{ZKB},  where
fluctuations in order to detect  the capacity regime are measured, can
be explained in the context of our simple, rule based model.

As the next issue,   we show how  far capacity can be  enhanced by
changing characteristics of  the vehicles or  of the driving behavior.
We  analyze the    influence of  a  ``cruise control'',   of  quicker
acceleration,   of  braking ``to  the  point'',  and of  a  better car
following behavior in the ``dead zone'' where neither acceleration nor
deceleration   are necessary.   Technically,   we  defined   different
``randomization probabilities'' $p_{acc}$,  $p_{sld}$,     $p_{free}$,
$p_{ptn}$  and $p_{ptn\_max}$  for   acceleration noise, noise  during
slowing  down, noise at free driving,  noise for platoon behavior, and
noise for platoon behavior   at   maximum speed, respectively.     The
definitions will be in a   way that $p_{acc}   = p_{sld} = p_{free}  =
p_{ptn}  =  p_{ptn\_max}  = 0.5$  reduces  to  the  ``old'' model with
$p_{general} = 0.5$.  These new noise  parameters were integrated into
the velocity update in the following way:\bi

\item
{\it acceleration:} If  the gap (see  above) to the next vehicle ahead
is large enough ($v \le gap - 1$) {\it and\/}  maximum velocity is not
yet reached ($v \le v_{max} - 1$), then accelerate with probability $1
-   p_{acc}$ by one:  $v  \to v + 1   -  \gamma(p_{acc})$.  (As above,
$\gamma(p)$ is one  with probability $p$ and  zero  elsewhen.)  A high
value of $p_{acc}$  therefore means that  vehicles have a tendency not
to accelerate even if they could.

\item
{\it slowing down:}  If  the next  car ahead is   too close ($gap  \le
v-1$),  then reduce the velocity,   with a probability of $p_{sld}$  to
overreact: $v \to gap - \gamma(p_{sld})$.

\item
{\it free driving:}  If the car has  maximum speed ($v = v_{max}$) and
drives freely ($gap \ge v_{max} + 1$), then introduce with probability
$p_{free}$ a fluctuation: $v \to v - \gamma(p_{free})$.

\item
{\it driving in a platoon at maximum speed:} If the car has maximum
speed ($v = v_{max}$) but is driving in a platoon ($v = gap$), then
slow down with probability $p_{ptn\_max}$ ($v \to v -
\gamma(p_{ptn\_max})$).

\item
{\it driving in a platoon with reduced speed:} If the car is driving
in a platoon ($v = gap$) with lower than maximum speed ($v \le
v_{max}-1$), then reduce speed with probability $p_{ptn}$ by one: $v
\to v - \gamma(p_{ptn})$.

\ei
As usual, we performed after this velocity update the vehicle
propagation.

Fig.~\ref{politics} contains the averaged fundamental diagrams when\bi

\item
$p_{acc}$ is reduced from $0.5$ to $0.005$ (better acceleration), or
when

\item
$p_{sld}$ is reduced from $0.5$ to $0.005$ (reduced overreaction for
slowing down), or when

\item
$p_{free}$ is reduced from $0.5$ to $0.005$ (reduced fluctuations at
free driving: ``cruise control''), or when

\item
$p_{ptn}$ and $p_{ptn\_max}$ are reduced from $0.5$ to $0.005$
(reduced fluctuations during platoon driving).

\ei
Reducing $p_{ptn\_max}$ in addition to $p_{free}$ (a case closer
inspected in a paper in preparation) gave no visible difference.

A cruise control  gives 0.324  vehicles  per iteration (2\%),   better
braking  gives 0.327 vehicles  per iteration (2\%), and better platoon
behavior  leads to an increase to  0.380 vehicles per iteration (about
20\%).   But the  remarkable  result of these  simulations  is that an
enhancement of  acceleration  ($p_{acc}$  reduced)  nearly doubles the
throughput     from 0.318     to   0.623    vehicles  per    iteration
(cf.~\cite{Piper} for  a    similar prediction).  In    addition,  the
space-time diagram for this system near capacity (Fig.~\ref{evol.acc})
looks qualitatively different from all  the others, which look, at
least at this resolution, similar to Fig.~\ref{evol.std}.

Although these results might need confirmation by more extensive
models, they present interesting information which should be carefully
evaluated before starting to equip vehicles with technical devices
such as automatic, radar-based car-following devices.

\sec Outlook

Sand falling down a very narrow glass tube shows clogging as well as
density waves reminiscent of traffic jams~\cite{Poeschl,%
LeibL}.  Quite general models for one-dimensional driven transport
systems indicate
self-organization~(\cite{Krug};
\footnote{
J.M.~Carlson, E.R.~Grannan, C.~Singh, G.H.~Swindle, Fluctuations in
Self-Organizing Systems, preprint 1993}) of the state of maximum
throughput.  Indeed, we found in our simulations that the outflow of a
jam self-organizes into maximum throughput, but we still have to check
the range of the validity of this result, especially to what extent it is
robust under different conditions such as different mixtures of vehicles.

Some of this work predicts in addition that this state (of maximum
throughput) should be ``critical'', pointing to a ``self-organizing
critical state''~\cite{Bak}.  This would support arguments of a phase
transition from laminar to turbulent traffic flow~\cite{Kue,Persaud}.
We have measured the life-times of the traffic jams in our model, but
the results indicate that criticality at maximum flow in these
observables is destroyed by the amount of noise introduced by the
fluctuations.

Along these lines, we hope to be able to relate the theory of traffic
flow to the general theory of driven diffusive systems.

\sec
Summary

We have introduced a low-fidelity microscopic traffic simulation model
based on very simple rules describing driving behavior.  Nevertheless,
this model proves to yield astonishingly realistic behavior.  We have
implemented two different codings of the model on six different computer
architectures, showing that, due to its simplicity, the model runs
efficient on these machines.  It can therefore be used for the
analysis of statistical properties of the simulated traffic, and in
addition, it is a candidate for being used to develop dynamic routing
methods.

Simulations have not only shown that the overall form of the
fundamental diagram as well as the formation of start-stop-waves are
robust phenomena in the sense that they persist under considerable
changes of the free parameters (see~\cite{NaS93} for some more
results), but it is even possible to use changed rules to predict
changes in traffic patterns.  The most remarkable of these predictions
is that the most efficient way to enhance throughput is the increase
of the acceleration capabilities of each individual vehicle.  Such
changes seem to be able to more than double capacity, whereas with all
other changes we never were able to gain more than 20\%.

\sec Acknowledgments

I thank C.~Barrett and S.~Rasmussen for discussions and the latter in
addition for revising the manuscript, T.~Pfenning for giving me
further insight into the details of coupled workstations, and SFI as
well as the TRANSIMS group at LANL for hospitality.  The NEC-SX3/11 of
the Regionales Rechenzentrum K\"oln provided the computing time for
Fig.~\ref{politics}.  This work has been supported by the
``Graduiertenkolleg Scientific Computing'' of the Land
Nordrhein-Westfalen and of the Bundesrepublik Deutschland and by the
TRANSIMS group of A-DO/SA (LANL).

\vfill\eject\noindent{\bf Table Captions}\bigskip

\noindent{TABLE~1:}
Computing   speed  of   different  algorithms  on  different  computer
architectures.  ``Vehicle-oriented'', ``intermediate'', and
``multi-lane'' mean the corresponding algorithms described in the
text; ``single-bit'' refers to cellular automaton techniques with single bit
coding.  For each machine and algorithm, the first table entry gives
the time each computer needed to simulate a system of
size~$10,000~km$.  From this figure, we derive the second and third
entries, which are real time limits in $km$ and in $vehicle \, sec /
sec$.  For further details see text.  The entries for multi-lane
traffic are due to M.~Rickert (personal communication).

\vfill\eject\noindent{\bf Figure Captions}
\newcounter{figno}
\setcounter{figno}{0}
\def\fig#1{\bigskip\noindent\refstepcounter{figno}{FIGURE~\thefigno:
}\label{#1}}

\fig{evol.std}
Evolution of the model from random initial conditions.  Each black
pixel represents a vehicle.  Space direction is horizontal, time is
pointing downwards, vehicles move to the right.  The simulation was of
a system of size $L=10000$ with density $\rho \approx \rho(q_{max})
\approx 0.08$; the figure shows the first 1000~iterations in a window
of $l=1000$.

\fig{fdiag}
{\it Top:\/ }Fundamental  diagram  of  the  model  (throughput  versus
density).  Triangles: Averages over short  times  (200~iterations) in
a sufficiently large system ($L = 10,000$).
Solid line: Long time averages ($10^6$~iterations) in a large system
($L = 10,000$).  Dashed line: Long
time averages ($10^6$~iterations) for a small system~$L=100$.
{\it Bottom:\/} Fluctuations of local velocity (see text) vs.\
density.

\fig{politics}
Throughput versus density for different sets of parameters (see text).
The legends gives the parameter(s) which is/are reduced versus the
``standard'' model.
Note the high increase in possible throughput when $p_{acc}$ is
reduced to 0.005 (vehicles accelerate more quickly).

\fig{evol.acc}
Evolution of the model from random initial conditions for
a reduced $p_{acc} = 0.005$.  Apart from that the figure is the same
as Fig.~\protect\ref{evol.std}.

\vfill\eject
\hoffset-1.6in \textwidth8.5in \advance\textwidth by-1in
\pagestyle{empty}
\offinterlineskip
\halign{&\tabskip0.5em\vbox to9pt{}\strut#\hfil&#\tabskip-0.5em\cr
\noalign{\noindent
TABLE~1: Computing speed of different algorithms on different
computer architectures\vskip1mm}
\noalign{\hrule}
Algorithm && Language
&& Sparc10
	&& PVM
		&& SX-3/11
			&& GCel-3
				&& CM-5
					&& iPSC \cr
   &&
&&
        && (5 $\times$ Sparc10)
         	&& 1 node${}^{v)}$
			&& 1024 nodes
				&& 32 nodes${}^{v)}$
					&& 32 nodes \cr
\noalign{\hrule}
\noalign{\vskip0.1pt{}\hrule}
Vehicle-or. && F77
&& 0.43~$sec$ && && && && &&\cr
  &&
&& (23,000~$km$) && && && && &&\cr
  &&
&& (0.31$\cdot 10^6~v\,s/s$)  && && && && && \cr
\noalign{\hrule}
Single-bit && F77
&& 0.33~$sec$
	&& 0.07~$sec$
		&& 0.0025~$sec$
			&& 0.013~$sec$
				&& 0.0077~$sec$
					&& 0.016~$sec$ \cr
  &&
&& (30,000~$km$)
	&& (140,000~$km$)
		&& (4,000,000~$km$)
			&& (750,000~$km$)
				&& (1,300,000~$km$)${}^{1)}$
					&& (630,000~$km$) \cr
  &&
&& (0.4$\cdot 10^6~v\,s/s$)
	&& (1.9$\cdot 10^6~v\,s/s$)
		&& (53$\cdot 10^6~v\,s/s$)
			&& (10$\cdot 10^6~v\,s/s$)
				&& (17$\cdot 10^6~v\,s/s$)
					&& (8.0$\cdot 10^6~v\,s/s$)
\cr
\noalign{\hrule}
Intermed. && C
&& 0.71~$sec$
	&& 0.15~$sec$
		&& 0.48~$sec$
			&& 0.011~$sec$
				&& 0.045~$sec$
					&& 0.038~$sec$ \cr
  &&
&& (14,000~$km$)
	&& (65,000~$km$)
		&& (21,000~$km$)
			&& (900,000~$km$)
				&& (220,000~$km$)${}^{2)}$
					&& (260,000~$km$) \cr
  &&
&& (0.19$\cdot 10^6~v\,s/s$)
	&& (0.87$\cdot 10^6~v\,s/s$)
		&& (0.28$\cdot 10^6~v\,s/s$)
			&& (12$\cdot 10^6~v\,s/s$)
				&& (2.9$\cdot 10^6~v\,s/s$) \
					&& (3.5$\cdot 10^6~v\,s/s$) \cr
\noalign{\hrule}
Intermed. && F77
&& 0.91~$sec$ && && && &&  &&  \cr
  &&
&& (11,000~$km$) && && && &&  &&  \cr
  &&
&& (0.15$\cdot 10^6~v\,s/s$)  && && && && && \cr
\noalign{\hrule}
Multi-lane && F77
&& 1.75~$sec$   &&    &&    &&    &&    && \cr
  &&
&& (5,700~$km$)   &&    &&    &&    &&    && \cr
  &&
&& (0.076$\cdot 10^6~v\,s/s$) && && && && && \cr
\noalign{\hrule}
}
\noindent ${}^{v)}$ Node(s) has/have vector units (SIMD instruction set)

\noindent ${}^{1)}$ using data parallel Fortran (CMF)

\noindent ${}^{2)}$ using message passing (CMMD)

\end{document}